\documentclass[prb,reprint,amssymb]{revtex4-2}

\usepackage{amsmath}  
\usepackage{amsfonts}
\usepackage{graphicx}
\usepackage{hyperref}
\hypersetup{
	colorlinks=true,
	citecolor=blue,
	linkcolor=red,
	urlcolor = blue
}
\usepackage{bm}

\begin{document}
	

\title{Ultrafast optical excitation of magnons in 2D antiferromagnetic semiconductors via spin torque mediated by  unbound electron-hole pairs and excitons: Signatures in magnonic charge pumping}
    
\author{Jalil Varela-Manjarres}
\author{Yafei Ren}
\author{Branislav K. Nikoli\'c}
    \email{bnikolic@udel.edu}
\affiliation{Department of Physics and Astronomy, University of Delaware, Newark, DE 19716, USA}
	
		
\begin{abstract}
Recent experiments observing how femtosecond laser pulse (fsLP) excites magnons in two-dimensional (2D) antiferromagnetic (AF) semiconductors---such as CrSBr, 
NiPS$_3$, and MnPS$_3$, or their van der Waals heterostructures---suggest an important role played by excitons. However, microscopic details of such an effect remain obscure, as resonant coupling of magnons, living in the sub-meV energy range, to excitons, living in the  \mbox{$\sim 1$ eV} range, can hardly be operative. Here, we develop a quantum transport theory of this effect, in which the time-dependent nonequilibrium Green's function  (TDNEGF) for electrons driven by fsLP is coupled self-consistently to the Landau-Lifshitz-Gilbert (LLG) equation describing classical dynamics of localized magnetic moments (LMMs) residing on magnetic atoms of 2D AF semiconductors. This theory explains how fsLP, of central frequency {\em above} the semiconductor gap, generates a photocurrent that becomes spin-polarized due to the background of LMMs, which, in turn, exerts spin-transfer torque (STT) onto LMMs as a genuinely nonequilibrium spintronic mechanism. By analyzing the collective motion of LMMs via the windowed Fast Fourier transform (FFT), we extract frequencies of excited magnons, as well as their lifetime governed by {\em nonlocal} damping with the LLG equation due to (explicitly included via TDNEGF) electronic bath. The TDNEGF part of the loop can also include excitons, which is achieved in the present study by utilizing the off-diagonal elements of the nonequilibrium density matrix as the mean-field treatment of Coulomb-binding of electron-hole pairs into excitons. Finally, our theory makes it possible to predict  that excited magnons will {\em pump} time-dependent charge currents into the attached electrodes, or locally within the 2D AF semiconductor, thereby emitting electromagnetic radiation. The windowed FFT of these two signals contains imprints of excited magnons, as well as the possible presence of excitons, which could be exploited as a novel probe in future experiments.  
\end{abstract}
	
\maketitle
	
\section{Introduction}

The advent of two-dimensional (2D) magnetic materials~\cite{Gibertini2019,Gish2024}---such as 2D antiferromagnetic (AF) semiconductors CrSBr~\cite{Bae2022,Diederich2022,Diederich2025,Brennan2024,wang2023magnetically, dirnberger2023magneto,Sun2024,Datta2025}, NiPS$_3$~\cite{Belvin2021},  MnPS$_3$~\cite{Wang2023} or their van der Waals (vdW) heterostructures~\cite{Onga2020,Zhang2020a}---has made possible recent experiments  observing how femtosecond laser pulse (fsLP)  excites magnons with possible exciton mediation. Excitons are  quasiparticles composed of an electron-hole pair bound by Coulomb interaction~\cite{Murakami2020,Perfetto2022,Perfetto2023,Semina2025,Cistaro2022,Meineke2024}, where imaging~\cite{Dong2021,Man2021} shows particularly large spatial extent in the case of CrSBr~\cite{Smiertka2026,Liebich2025,Shao2025,Brennan2024}. From a fundamental viewpoint, these experiments  rekindle~\cite{Bossini2021} interest in exciton-magnon coupling  that was dormant for many decades---for example, studies~\cite{Tanabe1965,Sell1967,Freeman1968,Parkinson1969,Gochev1975,Loginov1981,Gorbach1988} of such coupling in the 1960s and 1970s were largely motivated by absorption spectra of particular three-dimensional AF insulators like MnF$_2$. In the contemporary context of 2D magnetic materials, coupling of magnetism and excitons has revealed novel quantum many-body effects, such as:  magnon-magnon interactions~\cite{Zhitomirsky2013,Zheng2023} dressed by excitons~\cite{Diederich2025}; magnon-mediated exciton-exciton interactions~\cite{Datta2025,Johansen2019}; ultrafast exciton relaxation assisted by paramagnons~\cite{Meineke2024};  and  controlling quantum confinement (such as to a monolayer) or internal structure (i.e., the strength of Coulomb interaction binding) of excitons via AF order~\cite{Liebich2025,Shao2025}. 

As regards applications, in magnonics~\cite{Flebus2024} for classical information processing, there is a considerable effort to excite  coherent magnons~\cite{Pirro2021,De2024} by ultrafast light, whose frequencies are as high as possible and wavelengths as short as possible~\cite{Hortensius2021,Wang2023}. This is because magnons with a wavelength of \mbox{$\lesssim 100$ nm} would enable the miniaturization of envisaged magnonic devices down to the nanoscale~\cite{Chumak2015,Chumak2022}. In contrast, oscillating magnetic fields (supplied via microstrip lines or coplanar waveguides), as a standard tool to excite magnons in ferromagnets, are impractical for AF materials. Other schemes demonstrating excitation of AF magnons, such as by injecting  current~\cite{Lebrun2018}, lead to diffusive propagation of incoherent~\cite{De2024} (i.e., at many frequencies) magnons. Furthermore, for quantum information processing,  exciton-magnon coupling offers potential for 
transduction~\cite{Hisatomi2016,Lauk2020,Gish2024} of quantum information from qubits to microwaves that would excite magnons, and then from them, via excitons, further transduction to optical photons. In turn, photons can transfer quantum information over long distances via optical fibers~\cite{Lauk2020}.

\begin{figure}
		\centering
		\includegraphics[scale=0.7]{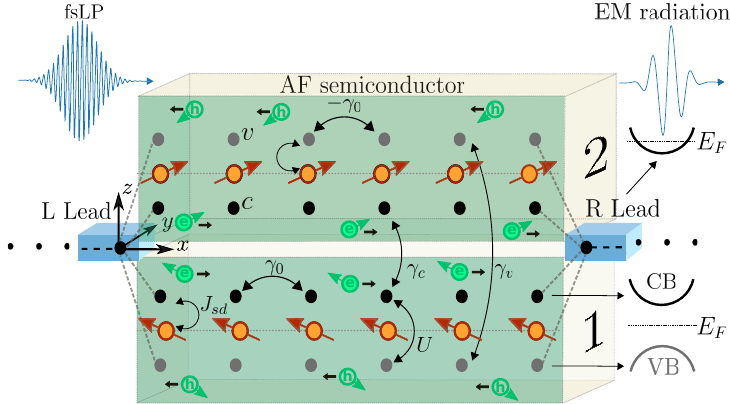}
		\caption{Schematic view of a two-terminal setup for TDNEGF+LLG+EX [Fig.~\ref{fig:fig1}] calculations of ultrafast dynamics of photoexcited electrons by {\em above} CB-VB gap fsLP within 2D AF semiconductor.  The central active region (CAR)   consists of two monolayers described by the top and bottom TB chains,  as inspired by the quasi-1D structure of CrSBr~\cite{Klein2023,Klein2024,Scheie2022,Bianchi2023,Esteras2022,Cui2025},  hosting two orbitals, $c$(onduction) and $v$(alence), per site $i$. Electron spin densities (green arrows), $\langle \hat{\mathbf{s}}_{i_c}\rangle (t)$ and $\langle \hat{\mathbf{s}}_{i_v}\rangle (t)$, interact via Kondo exchange [Eqs.~\eqref{eq:hamilc} and \eqref{eq:hamilq}] with LMM  at the same site described by the classical vector (red arrow) $\mathbf{M}_i(t)$ obeying the LLG Eq.~\eqref{eq:llg}. Note that LMMs are canted by applying an external magnetic field, as also used experimentally~\cite{Bae2022,Diederich2022,Diederich2025}, thereby introducing {\em noncollinearity} of LMMs between the two layers.  Electrons on $c$ and $v$ orbitals at the site $i$ interact via inter-orbital local Coulomb interaction of strength $U$ [Eq.~\eqref{eq:coulomb}], which, when turned on $U >0$, binds photoexcited electrons and holes into excitons~\cite{Murakami2020,Cistaro2022,Perfetto2022,Perfetto2023}.  Both chains are attached to semi-infinite ideal NM leads modeled also by 1D TB chains but without any interactions. Via such leads, any spin or charge current pumped within CAR by fsLP, or by excited magnons~\cite{Suresh2020,Evelt2017,Ciccarelli2015,Kapelrud2013} persisting after fsLP ceases, is drained [Fig.~\ref{fig:fig4}] toward macroscopic reservoirs kept at the same Fermi energy $E_F$ (i.e.,  no bias voltage is applied between the leads). We also compute EM radiation, emitted by pumped local charge currents within the CAR, and analyze its frequency content [Fig.~\ref{fig:fig3}].}
		\label{fig:fig0}
\end{figure}

\begin{figure}
	\centering
	\includegraphics[scale=0.6]{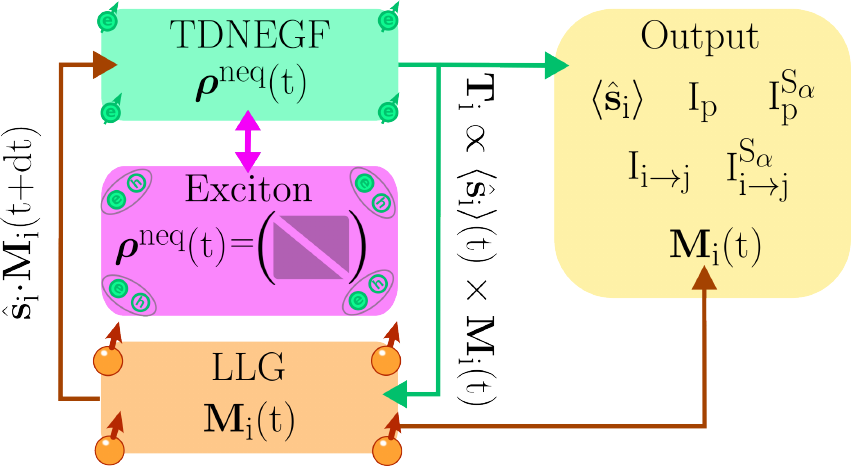}
	\caption{Flowchart of TDNEGF+LLG+EX self-consistent loop combining TDNEGF~\cite{Gaury2014,Popescu2016,stefanucci2025} (green box) computation [Eq.~\eqref{eq:rhoneq}] of time-dependent nonequilibrium density matrix ${\bm \rho}^{\rm neq}(t)$ with the LLG equation~\cite{Evans2014,Teuling2025} updating (orange box)  LMMs $\mathbf{M}_i(t)$. For this study, to the previously   developed~\cite{Petrovic2018,Bajpai2019a} TDNEGF+LLG scheme, we add  a computation (magenta box) employing~\cite{Cistaro2022} all  off-diagonal elements of ${\bm \rho}^{\rm neq}(t)$  to describe  the binding of photoexcited  conduction-band electrons and valence-band holes into excitons.  The loop employs  time step \mbox{$\delta t=0.1$ fs} in both quantum (as required for numerical stability of TDNEGF calculations~\cite{Gaury2014,Popescu2016,Petrovic2018}) and classical LLG calculations. After each time step, we obtain time-dependent observables  listed in the yellow box. In particular, STT is constructed from  the expectation value of electron spin $\langle \hat{\mathbf{s}}_i\rangle (t)$ [Eq.~\eqref{eq:spin}] and $\mathbf{M}_i(t)$   via Eq.~\eqref{eq:stt}.}
	\label{fig:fig1}
\end{figure}

 The possibility of precise tuning of the central frequency of fsLP around subgap electronic states (like exciton or on-site $d$-$d$ transition~\cite{Afanasiev2021,Allington2025}) and thereby  achieving control of excited magnons (such as dependence on light polarization~\cite{Allington2025}) emphasizes the {\em key role}~\cite{ReyesOsorio2025,KorffSchmising2024} played by photoexcited electrons as mediators of magnon excitation. Thus, electrons capable of responding  fast to fsLP {\em must be} explicitly included in any microscopic theory~\cite{Seifert2022a}. Their inclusion displaces often invoked~\cite{Satoh2010,Tzschaschel2017,Rongione2023,Seifert2019}, via intuitive reasoning, direct coupling of light to local magnetization of gapped materials---this is typically a negligible  effect when examined via first-principles calculations~\cite{Chen2019a}. However, the first-principles-derived Hamiltonians of exciton-magnon interactions are currently lacking~\cite{Kudlis2023}.  Furthermore, extraction of magnon and exciton spectra from a single first-principles framework, such as GW methodology~\cite{Olsen2021}, applied to examples of  2D AF semiconductors has revealed  their vastly different energy scales~\cite{Olsen2021}. This  means that direct (or resonant) exciton-magnon coupling is highly unlikely~\cite{Yuan2021}. One could still try to devise a Hamiltonian where off-resonant coupling emerges, but this is rarely considered as it requires special conditions~\cite{Yuan2021}. 

Lacking such inputs and microscopic theory based on them,  the magnon excitation aspect of recent experiments has usually been explained phenomenologically~\cite{Bae2022,Diederich2022,Diederich2025,Zhang2020a,Sun2024}, that is, by invoking an  ``impulsive perturbation''~\cite{Zhang2020a} of unknown origin that is introduced into the Landau-Lifshitz-Gilbert (LLG) equation~\cite{Evans2014} for localized magnetic moments (LMMs)   viewed as classical (unit) vectors $\mathbf{M}_i(t)$ [Fig.~\ref{fig:fig0}]. The agreement between such phenomenological models and experimentally excited magnon spectra implies that magnons can be approximately treated as classical spin waves~\cite{Teuling2025} describable by the LLG equation, despite challenges~\cite{GarciaGaitan2024,GarciaGaitan2025,Cichutek2025,Koerber2025} that AF materials can pose, especially~\cite{GarciaGaitan2024} those hosting LMMs based on spin $S=1/2$ or  $S=1$, for such an equation. In the case of 2D AF semiconductors with Cr as the source of LMM, the spin value is $S=3/2$ (e.g., in CrSBr, each Cr atom carries a net magnetic dipole moment of $3.03$ Bohr magneton  $\mu_B$, which agrees well with $S = 3/2$ for Cr$^{3+}$, while S and Br atoms are also slightly magnetized with \mbox{$-0.27$ $\mu_B$} and \mbox{$-0.08$ $\mu_B$}, respectively), so that a dissipative environment can induce transition~\cite{GarciaGaitan2024} from quantum to classical (i.e., LLG) dynamics.    We recall that in the LLG description,  magnons~\cite{Teuling2025,Zheng2023} emerge as collective excitations above a magnetically ordered ground state in which  $\mathbf{M}_i(t)$ precess around a direction specified by magnetic anisotropy and/or an externally applied magnetic field. The phase of precession $\mathbf{M}_i(t)$  of adjacent vectors varies  harmonically in space over the magnon wavelength $\lambda_m$.  However, microscopic understanding is lacking regarding what provides the initial ``kick'' for LLG dynamics to be initiated.  Furthermore, excited magnons could generate additional charge currents~\cite{Evelt2017,Ciccarelli2015,Suresh2020,Kapelrud2013} carrying their signatures within spectral content (note that initially generated photocurrent has spectral content related to a much higher frequency of light~\cite{Suresh2023,GarciaGaitan2025a}). The properties and usage of such signatures as potential novel experimental probes remain largely unexplored.

This study also employs the LLG equation, but we introduce the ``kick'' {\em microscopically} by self-consistently coupling [Fig.~\ref{fig:fig1}] the LLG equation to time-dependent  quantum transport~\cite{Gaury2014,Popescu2016} of photoexcited electrons, as described by   
the time-dependent nonequilibrium Green's functions (TDNEGF) formalism~\cite{Gaury2014,Popescu2016,stefanucci2025}. Thus, our TDNEGF+LLG framework~\cite{Petrovic2018,Bajpai2019a,Suresh2020,Petrovic2021,Suresh2021}, or TDNEGF+LLG+EX when excitons are included [Fig.~\ref{fig:fig1}], makes it possible to rigorously introduce into the LLG equation the effects brought by electronic photocurrent~\cite{Bajpai2019}. For example, photocurrent will become spin-polarized as it propagates through the magnetically ordered background of LMMs. In addition, since LMMs within one layer of  AF CAR in Fig.~\ref{fig:fig0} will be spin-polarized in a direction that is  {\em noncollinear} to LMMs in the second layer, electrons hopping from one layer will exert spin-transfer torque (STT)~\cite{Nikolic2018,Belashchenko2019,Nunez2006} onto the LMMs of the other layer: 
\begin{equation}\label{eq:stt}
\mathbf{T}_i (t)  =  J_{K} \langle \hat{\mathbf{s}}_{i} \rangle(t) \times \mathbf{M}_i(t),
\end{equation}
where the nonequilibrium expectation value of the  spin operator of an electron on the site $i$ is given by 
\begin{equation}\label{eq:spinoperatorsitei}
\langle \hat{\mathbf{s}}_{i}\rangle (t)  =  \left[ \sum_{a=c,v} \langle \hat{\mathbf{s}}_{ia}  \rangle (t) \right],
\end{equation}
considering that each site hosts two orbitals. 
Here, the nonequilibrium spin densities summed over  are given by
\begin{equation}\label{eq:spin}
\langle \hat{\mathbf{s}}_{ia} \rangle (t) = \mathrm{Tr}_\mathrm{spin} [{\bm \rho}^{\rm neq}(t) \hat{\mathbf{s}}_{ia}],  
\end{equation}
 $J_K$ is the  Kondo exchange interaction between the spin of flowing electrons and LMMs~\cite{Cooper1967,Tsvelik2017};    $a=c,v$ labels each of two orbitals per site [Fig.~\ref{fig:fig0}] with $\hat{\mathbf{s}}_{ia}$ being the spin operator of an  electron on the site $i$ in orbital $a$;   ${\bm \rho}^{\rm neq}(t)$ is the time-dependent nonequilibrium density matrix~\cite{Gaury2014,Popescu2016,Bajpai2020}; and the trace 
$\mathrm{Tr}_\mathrm{spin} [\ldots]$ is over quantum states in the spin space only. Importantly, without {\em canting} LMMs in Fig.~\ref{fig:fig0} by the applied magnetic field, as an {\em essential} ingredient in experiments on CrSBr~\cite{Bae2022,Diederich2022,Diederich2025},  
STT would be zero  (or unobservably small if we include noncolinearity only due to inevitable thermal fluctuations~\cite{Ghosh2022,Suresh2023}), so no magnons would be detectable as reported in experiments~\cite{Bae2022,Diederich2022,Diederich2025}. Such microscopically computed STT is then sent into the LLG equation 
\begin{eqnarray}\label{eq:llg}
	\partial_t \mathbf{M}_i & = & -g_0 \mathbf{M}_i \times \mathbf{B}^\mathrm{eff}_i + \alpha_G \mathbf{M}_i \times \partial_t \mathbf{M}_i  + \frac{g_0}{\mu_M}\mathbf{T}_i. 
\end{eqnarray}
In turn, dynamics of $\mathbf{M}_i(t)$ modifies the quantum Hamiltonian [Eq.~\eqref{eq:hamilq}] of electrons within the loop in Fig.~\ref{fig:fig1}. This establishes self-consistency between TDNEGF and LLG calculations, as the initially 
photoexcited spin current, \mbox{$ I^{S_{\alpha}}  = \mathrm{Tr} [{\bm \rho}^{\rm neq}(t) \hat{I}^{S_{\alpha}}]$}, will be modified by the dynamics of $\mathbf{M}_i(t)$;   whereas their trajectories are, in turn, affected by updated spin current and STT $\mathbf{T}_i(t)$ exerted by it. In the LLG Eq.~\eqref{eq:llg},  $g_0$ is the gyromagnetic factor~\cite{Evans2014}; $\mu_M$ is the magnitude of LMMs~\cite{Evans2014}; \mbox{$\mathbf{B}^{\rm eff}_i = - \frac{1}{\mu_M} \partial \mathcal{H} /\partial \mathbf{M}_{i}$} is the effective magnetic field obtained from {\em classical} Hamiltonian $\mathcal{H}$ [Eq.~\eqref{eq:hamilc}]  for LMMs~\cite{Scheie2022,Esteras2022,Teuling2025};  Gilbert damping is chosen as \mbox{$\alpha_G=0.01$}, which is the typical value for Cr-based 2D magnets~\cite{Esteras2022};  and we use shorthand notation $\partial_t \equiv \partial/\partial t$.

The paper  is organized as follows. The quantum Hamiltonian for the electronic subsystem, $\hat H$, and the classical  one, $\mathcal{H}$, for the subsystem of LMMs are introduced in Sec.~\ref{sec:hamiltonians}. The TDNEGF+LLG+EX methodology, using these Hamiltonians as an input, is covered in Sec.~\ref{sec:tdnegf}, as well as visually in Fig.~\ref{fig:fig1}. Section~\ref{sec:jefimenko} explaines the algorithm for the computation of  electromagnetic (EM) radiation via the Jefimenko equations~\cite{Jefimenko1966,Griffiths1991,McDonald1997,Ridley2021,Kefayati2024,VarelaManjarres2024,Francesco2025}, which uses bond currents as an input  obtained  from time-dependent quantum transport simulations via the TDNEGF+LLG+EX framework. We conclude and provide an outlook in Sec.~\ref{sec:conclusions}. In addition, we provide Fig.~\ref{fig:figS1} in the 
Appendix---with parameters from Fig.~\ref{fig:fig2} being artificially switched off or substantially increased, which then helps to further elucidate its results---as well as  a movie as the Supplemental Material (SM)~\footnote[1]{See Supplemental Material at \url{https://wiki.physics.udel.edu/qttg/Publications} for a movie animating the dynamics of LMMs and STT  exerted on them, as well as charge currents pumped by dynamical LMMs (i.e., by magnons as a collective dynamics of many LMMs).} animating Figs.~\ref{fig:fig2} and ~\ref{fig:fig4}. 

\section{Models and Methods}\label{sec:methods}

\subsection{Quantum and classical Hamiltonians}\label{sec:hamiltonians}

Our TDNEGF+LLG+EX calculations are fully microscopic,  requiring {\em only} two Hamiltonians as an 
input. These two Hamiltonians describe bare degrees of freedom---quantum Hamiltonian, $\hat{H}(t)$,  for electrons; and the classical one, $\mathcal{H}(t)$, for LMMs. In this study,  we focus on essential features of bilayer  2D AF semiconductors~\cite{Wang2023,Belvin2021,Afanasiev2021} by constructing a model quantum Hamiltonian [Eq.~\eqref{eq:hamilq}] that captures two monolayers, a semiconducting bandgap for each of them, ferromagnetic (FM)  intralayer and AF interlayer ordering of LMMs, and possible Coulomb interaction effects~\cite{Murakami2020,Cistaro2022}  binding conduction-band electrons and valence-band holes into excitons. Accordingly, our simulation setup in Fig.~\ref{fig:fig0} consists of two one-dimensional (1D) tight-binding (TB) chains~\cite{Klein2023,Klein2024,Cui2025} with 
two electronic orbitals per site $i$, as well as one LMM per site $i$ described by the unit vector $\mathbf{M}_i(t)$. Our usage of 1D chains is inspired by CrSBr~\cite{Klein2023,Klein2024,Scheie2022,Bianchi2023,Esteras2022,Cui2025} as a highly anisotropic 2D  magnetic material~\cite{Gibertini2019,Gish2024}  formed~\cite{Klein2023,Klein2024,Cui2025} by 1D atomic chains with interlayer  AF ordering (A-type) and intralayer  FM  ordering. The Curie temperature  of FM ordering within a single monolayer is \mbox{$\simeq 150$ K}, while AF ordering of LMMs between adjacent layers has a 
N\'{e}el temperature of \mbox{$\simeq 130$ K}~\cite{Bianchi2023}. Note that angle-resolved photoemission spectroscopy reveals the complex  electronic band structure of CrSBr, whose accurate theoretical description requires first-principles methods, such as DFT+U~\cite{Cui2025} or self-consistent GW methodology~\cite{Bianchi2023,Datta2025,Smiertka2026}. 

 The classical Hamiltonian, entering into the  LLG Eq.~\eqref{eq:llg}, is given by 
\begin{eqnarray}\label{eq:hamilc}
		\mathcal{H}(t) & = & J^{\mathrm{AF}}\!\!\!\sum_{\langle  i\in 1, j\in 2 \rangle} \mathbf{M}_i \cdot \mathbf{M}_j
        - J^{\mathrm{FM}}\!\!\!\sum_{\langle i\in 1, j\in 1 \rangle} \mathbf{M}_i \cdot \mathbf{M}_j\nonumber\\ &-&J^{\rm FM}\sum_{\langle i\in 2, j\in 2 \rangle} \mathbf{M}_i \cdot \mathbf{M}_j -K_x\sum_{i\in 1,2}(\mathbf{M}_i\cdot \mathbf{e}_x)^2\nonumber\\
        &+& 
         K_z\sum_{i\in 1,2}(\mathbf{M}_i\cdot \mathbf{e}_z)^2- J_{K}\sum_{i\in 1,2}\mathbf{M}_i\cdot\sum_{a=c,v} \langle \mathbf{\hat{s}}^a_i \rangle. 
\end{eqnarray}
Such an effective classical Heisenberg Hamiltonian of LMMs  can be extracted from first-principles calculations~\cite{Szilva2023,Olsen2021,Durhuus2023} or by fitting magnon spectra from neutron scattering data~\cite{Scheie2022}. 
Here \mbox{$J^{\rm AF}=0.0195\ \mathrm{eV}$} and \mbox{$J^{\rm FM}=0.15\  \mathrm{eV} $} are AF and FM exchange couplings, respectively; $K_x=0.021\  \mathrm{eV} $ and \mbox{$K_z=0.057\  \mathrm{eV} $} specify the magnetic anisotropy along the $x$- and $z$-axis, respectively; \mbox{$J_{K}=0.01\ \mathrm{eV}$} is Kondo exchange interaction between spins of flowing electrons and LMMs~\cite{Cooper1967,Tsvelik2017}; $i \in 1$ ($i \in 2$) means that the site $i$ belongs to the first (second) layer in Fig.~\ref{fig:fig0}; and $\langle\ldots\rangle$ denotes  NN sites.   Note that
 $J^{\rm AF},\ J^{\rm FM},\ K_x,\ K_z $ 
 parameters were also employed to interpret experiments~\cite{Bae2022,Diederich2022,Diederich2025}, but we rescale their values  by a multiplicative factor $\times 10^3$ to make the total TDNEGF+LLG+EX   simulation time of \mbox{$\sim 10$ ps} duration. Nevertheless, even with this adjustment,  the characteristic energy scale of the AF background remains below electron kinetic energy,  $\gamma_0/J^{\rm{AF}}\simeq 10^2$. Such rescaling shifts the spectral content of pumped charge current and its radiation into the THz range in Figs.~\ref{fig:fig4} and ~\ref{fig:fig3}, while for, e.g.,  CrSBr the relevant range would be at GHz frequencies.  

The same Kondo exchange 
\mbox{$\hat{H}_{K}(t)= - J_{K}\sum_i\mathbf{M}_i(t)\cdot\sum_{a=c,v}\hat{\mathbf{s}}_{ia}$} enters into the quantum Hamiltonian of  electrons 
\begin{equation}\label{eq:hamilq}
		\hat{H}(t)  =   \hat{H}_{K}  + \hat{H}_{\mathrm{intra}}  + \hat{H}_{\mathrm{inter}} +\hat{H}_{\rm Coulomb}.
\end{equation}
Here $\hat{H}_{\mathrm{intra}}$ describes 1D TB chains within layer 1 or 2 in Fig.~\ref{fig:fig0}  
\begin{eqnarray}\label{eq:hamil_intra}
    &&\hat{H}_{\mathrm{intra}} = \Delta/2\sum_{i\in 1,2}(\hat{c}_{i}^{c \dagger}  \hat{c}^c_{i }- \hat{c}_{i}^{v \dagger}  \hat{c}^v_{i} )  \\
    && +  \gamma_0\sum_{\langle i\in 1 j\in 2 \rangle} e^{i\chi_{ij}(t)}( \hat{c}_{i}^{v \dagger}  \hat{c}^v_{j } - \hat{c}_{i}^{c\dagger}  \hat{c}^c_{j } ) - \gamma_{\mathrm{P}}(t)\sum_{i\in 1,2} \hat{c}_{i}^{c \dagger}  \hat{c}^v_{i}  + \mathrm{H.c}, \nonumber
\end{eqnarray}
where indices $c$ and $v$ stand for orbitals that give rise to the conduction and valence bands of each chain; \mbox{$\hat{c}_i^{a\dagger}=(\hat{c}_{i\uparrow}^{a\dagger} \  \ \hat{c}_{i\downarrow}^{a\dagger})$} is the row vector containing operators $\hat{c}_{i\sigma}^{a\dagger}$ creating  electrons with spin \mbox{$\sigma=\uparrow,\downarrow$} in orbital \mbox{$a=c,v$} hosted by site $i$; 
$\Delta$ is the onsite potential opening a bandgap  between the two bands $\Delta = 3\ \rm{eV}$; and 
  $\gamma_0 = 1\ \mathrm{eV}$ is the hopping parameter between the NN sites.  The spin density operator in Eqs.~\eqref{eq:spin} and ~\eqref{eq:hamilq} is given by $\mathbf{\hat{s}}_{ia}= \hat{c}_{i}^{a \dagger} \hat{\bm \sigma} \hat{c}^a_{i}$ for $a=c,v$ and  \mbox{${\bm  \sigma} = (\hat{\sigma}_x,\hat{\sigma}_y,\hat{\sigma}_z)$} being  the vector of the Pauli matrices. The fsLP is introduced in Eq.~\eqref{eq:hamil_intra} via the Peierls phase~\cite{Eckstein2020,Panati2003}, \mbox{$\chi_{ij}(t) =z_{\max } \exp [-\left(t-t_p\right)^2 /(2 \sigma_{\text {light }}^2)] \sin (\omega_0 t) $}, where the electric field of the pulse is $\mathbf{E}(t)=-\partial_t \mathbf{A}(t)$,    \mbox{$z_{\max}=eaA_{\rm max}/\hbar=0.1$} is a dimensionless parameter quantifying the intensity of the pulse using $A_{\rm max}$ as the amplitude of the vector potential, $\omega_0$ is the central frequency of fsLP,  and $a$ is the lattice spacing. Besides the Peierls phase, fsLP is additionally~\cite{Murakami2020}  introduced via hopping  $\gamma_{\mathrm{P}}(t) = \mathbf{d}\cdot \mathbf{E}(t)$ in Eq.~\eqref{eq:hamil_intra}. This term accounts for interband transitions  driven by the dipole interaction with the electric field~\cite{Murakami2020}, where $\mathbf{d}$ is the expectation value of the dipole operator, $\mathbf{d} = e\langle i,c| \mathbf{\hat{r}}|i,v\rangle$. The term $\hat{H}_{\mathrm{inter}}$ is given by 
\begin{equation}\label{eq:hamil_inter}
    \hat{H}_{\mathrm{inter}}(t) = -\sum_{\langle i \in 1,j \in 2\rangle} (\gamma_{c} \hat{c}_{i}^{c\dagger}  \hat{c}^{c}_{j}-\gamma_{v} \hat{c}_{i}^{v \dagger}  \hat{c}^{v}_{j}  +\mathrm{H.c.}),
\end{equation}
and it describes hopping with parameters $\gamma_{c}=0.5\gamma_0$ or $\gamma_{v}=0.5\gamma_0$ between $c$ or $v$ orbitals, respectively, located at NN sites of two different chains. Finally, the inter-orbital  and intralayer  Coulomb interaction~\cite{Murakami2020,Cistaro2022} 
\begin{equation}\label{eq:coulomb}
   \hat H_{\rm Coulomb}= U \sum_{i\in 1,2;\sigma;\sigma'} \hat{n}^c_{i,\sigma} \hat{n}^v_{i,\sigma' },
\end{equation}
 describes how two electrons on two different orbitals at the same site $i$ interact with each other. The same Hubbard term was employed in prior studies~\cite{Murakami2020,Cistaro2022} to describe exciton formation on the TB lattice. We decouple it via tMFT~\cite{Cistaro2022} (otherwise, its beyond-tMFT  treatment requires computationally much more expensive evaluation of Feynman diagrams on the Keldysh contour~\cite{Murakami2020}) as follows:
\begin{eqnarray} \label{eq:hamil_mfi}
\hat{n}^c_i \hat{n}^v_i &\rightarrow &\langle \hat{n}^c_i(t) \rangle  \hat{n}^v_i +  \langle \hat{n}^v_i(t) \rangle  \hat{n}^c_i\nonumber \\
&-&    \phi_i(t)  \hat{c}_i^{c \dagger}  \hat{c}^{v}_i
 - \phi_i^*(t)  \hat{c}_i^{v \dagger}  \hat{c}^{c}_i.
\end{eqnarray}
Here the first two terms correspond to Hartree and the latter two to Fock approximation, where the order parameter of the excitonic condensate~\cite{Johansen2019} is given by 
$\phi_i(t) = \langle \hat{c}_{i}^{v \dagger}(t)  \hat{c}^{c}_{i}(t) \rangle$. This procedure requires us to self-consistently compute $\phi_i(t)$, which we obtain from the off-diagonal elements of   ${\bm \rho}^{\rm neq}(t)$ as a part of TDNEGF calculations within  the self-consistent loop illustrated in Fig.~\ref{fig:fig1}. We note that in this approach we do not rely on the effective excitonic Hamiltonians written in terms of operators of composite bosons~\cite{Kudlis2023}, which can be cumbersome to derive~\cite{Combescot2002}. Instead, we use ``bare'' degrees of freedom, so that {\em both}  unbound electron-hole pairs and their binding into excitons by the Coulomb interaction are considered through the  electronic density matrix, as naturally included~\cite{Cistaro2022,Murakami2020,Perfetto2022,Perfetto2023} when using the nonequilibrium density matrix and/or TDNEGF formalism. 
 
The AF semiconductor CAR in Fig.~\ref{fig:fig0} is attached to  $L$ and  $R$ normal metal (NM) leads modeled as semi-infinite ideal 1D TB chains with one orbital per site. The chemical potential of the macroscopic reservoirs into which NM leads terminate is identical (i.e., no DC bias voltage is applied) and chosen as $\mu_L=\mu_R=E_F=0$. 

\subsection{TDNEGF+LLG+EX methodology}\label{sec:tdnegf}

The fundamental quantity of quantum statistical mechanics is the density matrix. The time-dependent one-particle nonequilibrium density matrix,  ${\bm \rho^{\rm neq}}(t) = \hbar \mathbf{G}^<(t,t)/i$, can be expressed in terms of the lesser Green's function (GF)  of TDNEGF formalism \mbox{$G^{<,\sigma\sigma'}_{ii'}(t,t')=\frac{i}{\hbar} \langle \hat{c}^\dagger_{i'\sigma'}(t') \hat{c}_{i\sigma}(t)\rangle_\mathrm{nes}$} ~\cite{Gaury2014} where  $\langle \ldots \rangle_\mathrm{nes}$ is the nonequilibrium statistical average~\cite{stefanucci2025}. We solve a matrix integro-differential equation~\cite{Popescu2016}  
\begin{equation}\label{eq:rhoneq}
		i\hbar \partial_t {\bm \rho}^{\rm neq} = [\mathbf{H}(t),{\bm \rho}^{\rm neq}] + i \sum_{ p=L,R} [{\bm \Pi}_p(t) + {\bm \Pi}_p^{\dagger}(t)],
\end{equation}
for the time evolution of ${\bm \rho}^{\rm neq}(t)$, where $\mathbf{H}(t)$ is the matrix representation of the quantum Hamiltonian of electrons. Equation~\eqref{eq:rhoneq} is an {\em exact} quantum master equation for the reduced density matrix of the AF CAR in Fig.~\ref{fig:fig1}  viewed as an open finite-size quantum system attached to macroscopic Fermi liquid reservoirs via semi-infinite NM leads.  The NM leads, {\em not exploited} in experiments~\cite{Bae2022,Diederich2022,wang2023magnetically, dirnberger2023magneto,Sun2024,Diederich2025,Datta2025,Belvin2021} thus far, are important technically  within TDNEGF calculations to introduce continuous energy spectrum and dissipation effects ~\cite{Joao2025,VarelaManjarres2023}, 
thereby guaranteeing that excited photocurrent and $\mathbf{M}_i(t)$ dynamics will eventually terminate~\cite{Suresh2023} after fsLP cases. Otherwise, in a closed quantum system~\cite{Ghosh2022,Kudlis2023} without a surrounding bath~\cite{Suresh2023} one would find a forever oscillating photocurrent (as it occurs, e.g., 
in first-principles time-dependent density functional theory calculations applied to closed quantum systems~\cite{Kefayati2024,Kefayati2025,Kefayati2025a}), which is unphysical. Furthermore, the NM leads  allow us to analyze properties of charge and spin currents drained into them. Such currents could offer a novel experimental probe of excitons, magnons, and their interactions, as we confirm in  Fig.~\ref{fig:fig4}. For this purpose, we use  ${\bm \Pi}_p(t)$ matrices
	\begin{equation}\label{eq:current}
		{\bm \Pi}_p(t) = \int_{0}^t \!\! dt_2\, [\mathbf{G}^>(t,t_2){\bm \Sigma}_p^<(t_2,t) 
		- \mathbf{G}^<(t,t_2){\bm \Sigma}_p^>(t_2,t) ],
	\end{equation} 
expressed in terms of the lesser and greater GFs~\cite{stefanucci2025} and the corresponding self-energies ${\bm \Sigma}_p^{>,<}(t,t')$~\cite{Popescu2016}, to obtain  time-dependent charge
\begin{eqnarray}\label{eq:charge_curr}
    I_p(t)  = \frac{e}{\hbar} \mathrm{Tr}\, [{\bm \Pi}_p(t)],
\end{eqnarray}
and spin
\begin{equation}\label{eq:spin_curr}
        I_p^{S_{\alpha}}(t) = \frac{e}{\hbar} \mathrm{Tr}\, [\hat{\sigma}_{\alpha}{\bm \Pi}_p(t)],
\end{equation}
currents outflowing into  $p = L,R$ NM leads.  	Since the  applied bias voltage between the left ($L$) and right ($R$) NM leads is identically zero in the setup of  Fig.~\ref{fig:fig0}, all computed currents $I_p(t)$ and $I_p^{S_{\alpha}}(t)$ are solely  due to {\em pumping} by nonperiodic~\cite{VarelaManjarres2024,Abbout2018,Petrovic2018,Petrovic2021} (instead of usually considered periodic~\cite{VarelaManjarres2024,Tserkovnyak2005,Suresh2020,Kapelrud2013}) time dependence of the CAR Hamiltonian. Note that we use the same units for charge and spin currents, defined as \mbox{$I_p = I_p^{\uparrow} + I_p^{\downarrow}$} and \mbox{$I_p^{S_{\alpha}} = I_p^{\uparrow} - I_p^{\downarrow}$}, in terms of spin-resolved charge currents $I_p^{\sigma}$. In our convention, positive current in the NM lead $p$ means charge or spin current is flowing out of that NM lead.

Let us recall that the problem of how STT  excites uniform motion of all LMMs [i.e., of their macrospin \mbox{$\mathbf{M}(t)=\sum_i \mathbf{M}_i(t)$}] vs. their nonuniform motion like magnons  was analyzed long ago~\cite{Brataas2006a} for conventional metallic ferromagnets, as well as  recently for AF insulators~\cite{Daniels2015,Suresh2021}, with a focus on threshold injected current value~\cite{Lebrun2018,Kajiwara2010} for magnons to occur. Our setup in  Fig.~\ref{fig:fig1} is different from those studies, as no current is injected from an external circuit. Instead, photocurrent is excited within the AF semiconductor CAR by fsLP.  In addition to the photocurrent of unbound electron-hole pairs, we also consider possibility of photocurrent of excitons due to Coulomb interaction [Eq.~\eqref{eq:coulomb}] binding conduction-band electrons with valence-band holes. To capture such binding, we employ  a time-dependent mean-field theory  (tMFT)~\cite{Cistaro2022,Murakami2020} of the inter-orbital  Coulomb interaction term [Eq.~\eqref{eq:coulomb}]. For this purpose, we  exploit~\cite{Cistaro2022} 
the off-diagonal elements of $\mathbf{\bm \rho}^{\rm neq}(t)$, that we naturally construct within the TDNEGF part~\cite{Popescu2016,Gaury2014} of the TDNEGF+LLG self-consistent loop~\cite{Petrovic2018,Bajpai2019a,Suresh2020,Petrovic2021,Suresh2021}. In other words, for the study conducted in Sec.~\ref{sec:results}, the previously developed~\cite{Petrovic2018,Bajpai2019a,Suresh2020,Petrovic2021,Suresh2021,qttgsoftware} TDNEGF+LLG is generalized into the  TDNEGF+LLG+EX framework by adding modeling of excitonic effects at tMFT level~\cite{Cistaro2022,Murakami2020}, as illustrated by magenta box in Fig.~\ref{fig:fig1}. 

\subsection{Electromagnetic radiation by bond charge currents from the  Jefimenko equations}\label{sec:jefimenko}

The electric field of EM radiation emitted into the FF region~\cite{McDonald1997,Kefayati2024} is calculated from the Jefimenko equations~\cite{Jefimenko1966}, reorganized~\cite{McDonald1997} to isolate the contribution in the FF region
\begin{widetext}
\begin{equation}\label{eq:efield}
\mathbf{E}_\mathrm{FF}(\mathbf{r}, t)=\frac{1}{4\pi \epsilon_0 c^2}\sum_{P_{ia \rightarrow jb}=1}^{N_B} \int_{P_{ia \rightarrow jb}}\bigg[ (\mathbf{r}-\mathbf{l})\frac{\partial_t I_{ia \rightarrow jb}(t_r)}{|\mathbf{r}-\mathbf{l}|^3}(\mathbf{r}-\mathbf{l})\cdot \mathbf{e}_x - \frac{\partial_t I_{ia \rightarrow jb}\left(t_r\right)}{|\mathbf{r}-\mathbf{l}|} \mathbf{e}_x \bigg ] d l.
\end{equation}
\end{widetext}
 Note that Jefimenko Eq.~\eqref{eq:efield} can be viewed~\cite{Griffiths1991} as proper (i.e., time-retarded)  time-dependent generalizations of the Coulomb  law. Here,  \mbox{$t_r \equiv t -|\mathbf{r}-\mathbf{l}|/c$} emphasizes retardation in the response time due to relativistic causality~\cite{Jefimenko1966,McDonald1997}. 
 Additionally, we adapt~\cite{Ridley2021,Suresh2023}  Eq.~\eqref{eq:efield}  to utilize time-dependent bond~\cite{Nikolic2006} charge currents as the source of EM radiation, $I_{ia \rightarrow jb}(t)$ [Eq.~\eqref{eq:bondcharge}]. They are the  counterpart (on the TB lattice) of local current density in continuous space. The bond currents $I_{ia\rightarrow jb}$ are assumed to be spatially homogeneous along the path $P_{ia\rightarrow jb}$ from orbital $a$ at site $i$ to orbital $b$ at site $j$~\cite{Nikolic2006,Petrovic2018,Ridley2021}, which is composed of a set of points $l \in P_{ia\rightarrow jb}$. Here $N_B$ is the number of bonds $ia\rightarrow jb$, and since we use $N=10$  black and gray sites [Fig.~\ref{fig:fig0}] in each layer of CrSBr,  $N_B=36$ in our calculations.  We obtain bond charge currents as 
\begin{equation}\label{eq:bondcharge}
	I_{ia\rightarrow jb}(t) = \frac{e\gamma}{i\hbar}\mathrm{Tr}_{\rm spin}\bigg[{\bm \rho}^{\rm neq}_{ia,jb}(t)\mathbf{H}_{jb,ia}(t) - {\bm \rho}^{\rm neq}_{jb,ia}(t)\mathbf{H}_{ia,jb}(t)\bigg].
\end{equation}
For this purpose, we isolate  $2 \times 2$ submatrices	${\bm \rho}^{\rm neq}_{ia,jb}(t)$  of ${\bm \rho}^{\rm neq}(t)$. Note that diagonal elements of ${\bm \rho}^{\rm neq}_{ij}(t)$ determine on-site nonequilibrium charge density, whose time dependence contributes to near-field  radiation~\cite{Ridley2021,Suresh2023,Kefayati2024}.

\section{Results and Discussion}\label{sec:results}

\begin{figure}
		\centering
		\includegraphics[scale=0.23]{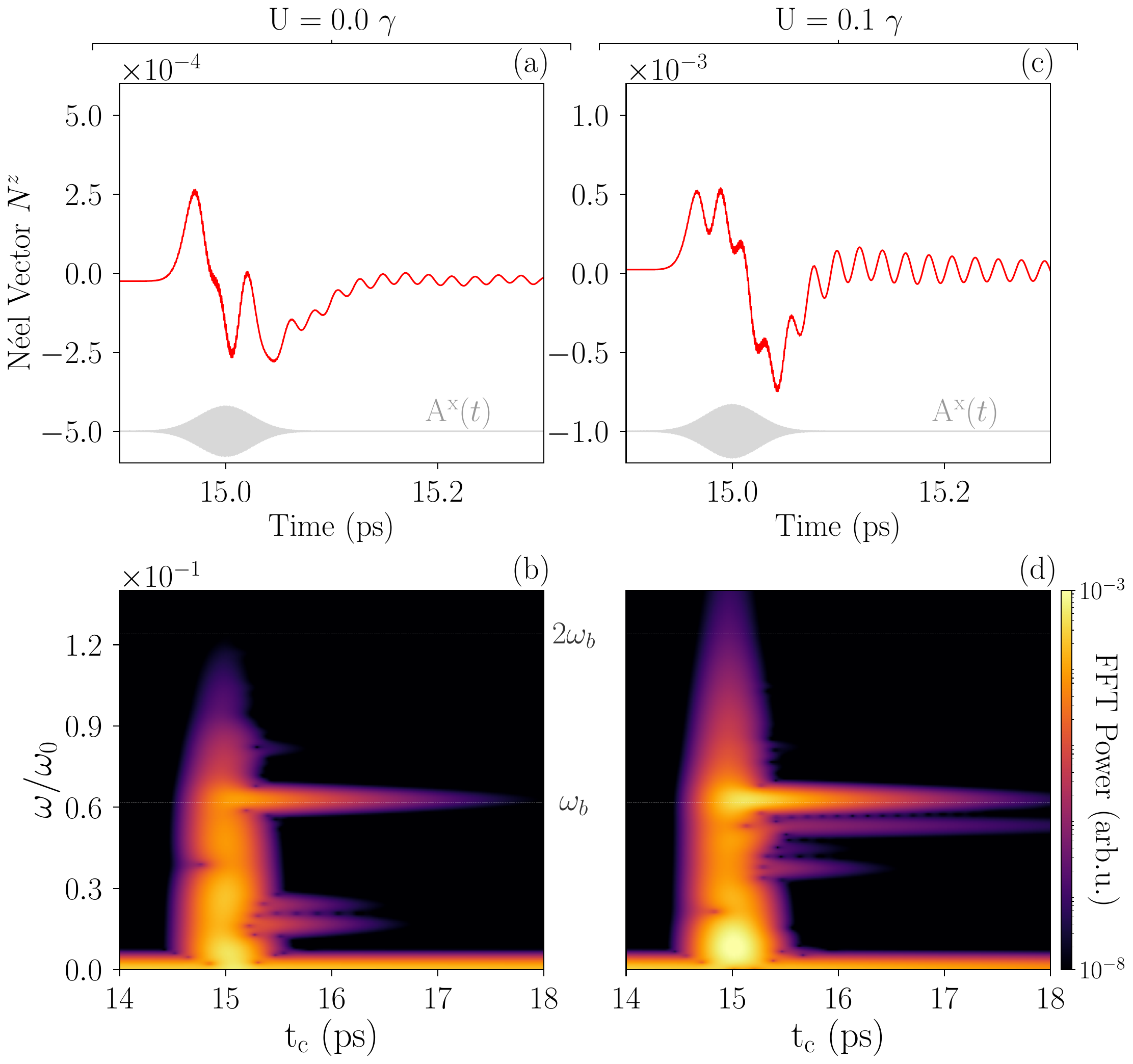}
		\caption{Time dependence of the N\'{e}el vector, defined [Eq.~\eqref{eq:Neel_vec}] for two monolayers of 2D AF semiconductor in Fig.~\ref{fig:fig0} in: (a) the absence of excitons ($U=0$); or (c) their presence induced by on-site Coulomb interaction~\cite{Murakami2020,Cistaro2022} $U=0.1\gamma$ [Eq.~\eqref{eq:coulomb}]. Panels (b) and (d)
    plot the corresponding power spectrum of windowed FFT~\cite{Press2007,Cohen2014}, $|N^z(t_c,\omega)|^2$, revealing frequencies and lifetimes of  magnons excited by STT from the photocurrent of unbound electron-hole pairs in (a) or those pairs and excitons acting together in (b), respectively. Here $t_c$ denotes the central time of the Gaussian window  \cite{Press2007,Cohen2014} used in FFT  [Eq.~\eqref{eq:FFT_window}]. Note that $\omega_b$ labels  the frequency of the ``bright magnon,'' which corresponds to the same type of magnon  observed in experiments of Refs.~\cite{Bae2022,Diederich2022,Diederich2025} on a bilayer of 2D AF semiconductor CrSBr. Gray curves on the bottom of panels (a) and (c) depict the vector potential $A^x(t)$ of fsLP.}
		\label{fig:fig2}
\end{figure}

We first recall  the definition of the N\'eel vector analyzed in recent experiments~\cite{Bae2022,Diederich2022,Diederich2025,Sun2024}
\begin{equation}\label{eq:Neel_vec}
    \mathbf{N}\equiv(N^x,N^y,N^z) = \frac{1}{2N}\sum_{i \in 1,j \in 2} \big(\mathbf{M}_{i}-\mathbf{M}_{j}\big),
\end{equation}
between two monolayers  of  AF semiconductor. Thus, in equilibrium and without any canting external magnetic field applied, we find $\mathbf{N}(t=0) \equiv  (2,0,0)$. Out of equilibrium, as initiated by fsLP, the N\'eel vector starts evolving  in time [Figs.~\ref{fig:fig2}(a) and \ref{fig:fig2}(c)]. Somewhat surprisingly and also  observed experimentally~\cite{Zhang2020a}, this evolution starts without substantial delay regarding fsLP, even though LMMs are slower than electrons~\cite{Suresh2023}. Because the magnon frequency spectrum encoded in $N^z(t)$ could be changing within different time frames, we apply windowed (or short-time) Fast Fourier transform (FFT)~\cite{Press2007,Cohen2014,Mina2022,Molinero2024,Jensen2024}.  For example, magnons get excited  around  $t_c \simeq 15\ \mathrm{ps}$ in Fig.~\ref{fig:fig2}(b), and subsequently they decay~\cite{Zhitomirsky2013,Flebus2024} because of Gilbert damping $\alpha_G$ in Eq.~\eqref{eq:llg}. Furthermore,  because of explicitly introduced electrons via TDNEGF calculations, additional {\em nonlocal} damping~\cite{Bajpai2019a,ReyesOsorio2024,Reyes2024,Sayad2015,Zhang2009,Yuan2016,Verba2018} is introduced into  Eq.~\eqref{eq:llg} via the STT term $\mathbf{T}_i(t)$. Due to such decay, excited magnons vanish at around  $t_c \simeq 18\ \mathrm{ps}$ in Fig.~\ref{fig:fig2}(b). 

\begin{figure}
		\centering
		\includegraphics[scale=0.23 
        ]{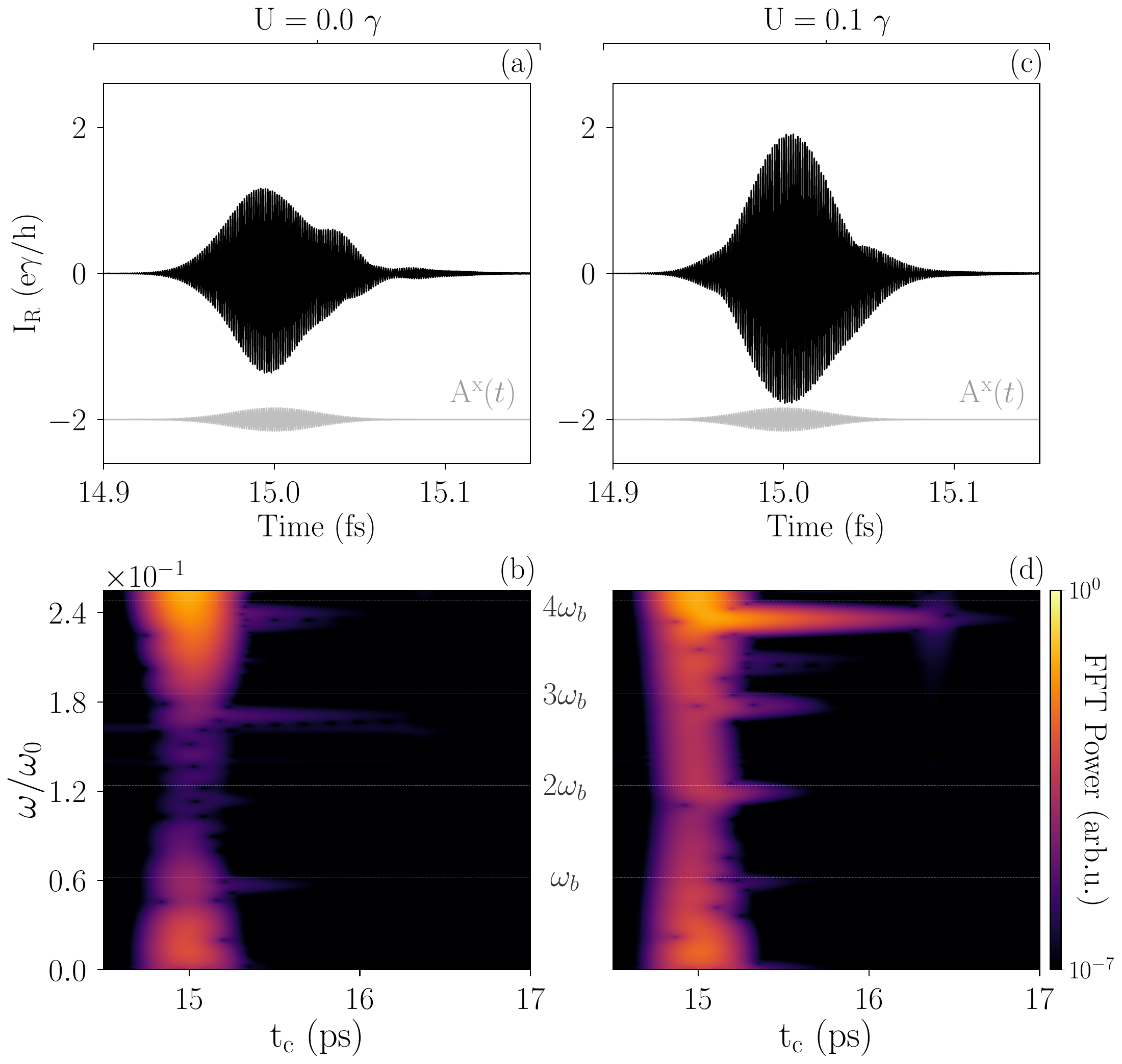}
		\caption{Time-dependence of charge current $I_R(t)$ flowing into the right NM lead in Fig.~\ref{fig:fig0} due to magnonic charge pumping~\cite{Suresh2020,Ciccarelli2015,Evelt2017,Kapelrud2013} by excited magnons from Fig.~\ref{fig:fig2} in: (a) the absence of excitons ($U=0$); or (c) their presence induced by on-site Coulomb interaction $U=0.1\gamma$. Panels (b) and (d) plot the corresponding power spectrum of windowed FFT~\cite{Press2007,Cohen2014}, $|I_R(t_c,\omega)|^2$. Gray curves on the bottom of panels (a) and (c) depict the vector potential  $A^x(t)$ of fsLP.}
		\label{fig:fig4}
\end{figure}

\begin{figure}[hbt!]
		\centering
		\includegraphics[scale=0.23]{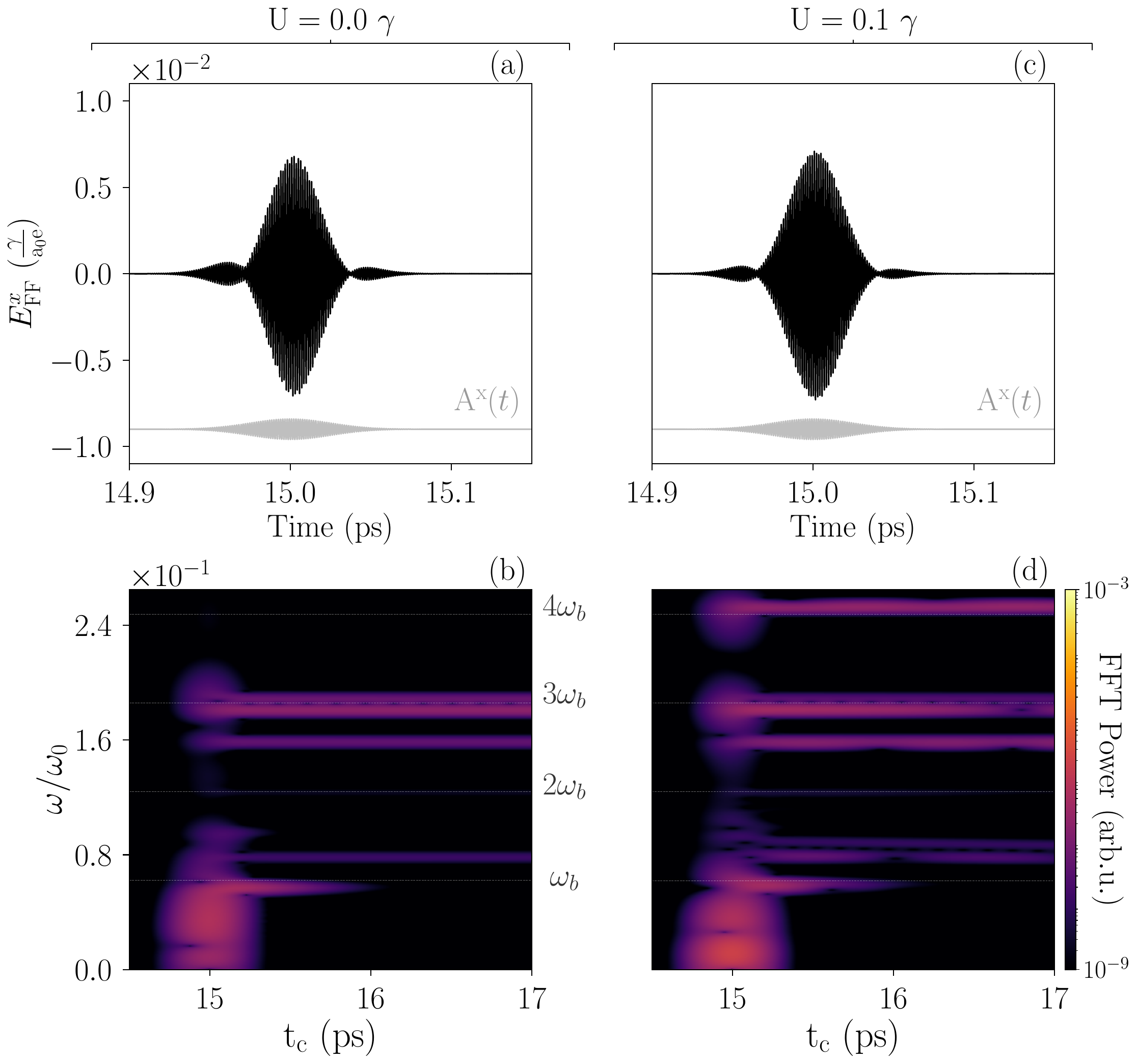}
		\caption{Time-dependence of the $E_\mathrm{FF}^x(t)$ component of the electric field of EM radiation emitted into  the FF region by excited ``bright magnon'' (from Fig.~\ref{fig:fig2}) pumping~\cite{Evelt2017,Suresh2020,Ciccarelli2015,Kapelrud2013} bond charge 
        currents [Eq.~\eqref{eq:bondcharge}] within the CAR of Fig.~\ref{fig:fig0} in: (a) the absence of excitons ($U=0$); or (c) their presence  induced by on-site Coulomb interaction $U=0.1\gamma$. Panels (b) and (d) plot the corresponding power spectrum of windowed FFT~\cite{Press2007,Cohen2014},  $|E_\mathrm{FF}^x(t_c,\omega)|^2$.  Gray curves on the bottom of panels (a) and (c) depict the  vector potential $A^x(t)$ of fsLP.}
		\label{fig:fig3}
\end{figure}

For such signals---appearing also in many other scientific disciplines (such as electroencephalography~\cite{Cohen2014} or speech analysis)---it is advantageous to perform windowed FFT over successive time intervals. For this purpose, we employ Gaussian as the  window function  
\begin{equation}\label{eq:FFT_window}
    X(t_c,\omega) = \frac{1}{\lambda\sigma(2\pi)^{3/2}}\int\limits_{0}^{\infty}\!\! dt \,X(t)e^{i\omega t}e^{-(t-t_c)^2/2(\lambda\sigma)^2},
\end{equation}
which makes it possible to extract time-frequency content from signals whose oscillations are localized in a finite  time frame.  Here, $t_c$, the centroid of the Gaussian, serves as the abscissa of panels (b) and (d) within each of Figs.~\ref{fig:fig2}--\ref{fig:fig3} analyzing $X(t)=N^z(t),I_R(t),E^x_{\mathrm{FF}}(t)$ as the signal,  respectively. Here,  $\sigma$ specifies the width of the Gaussian and  $\lambda$ is a parameter controlling the resolution. For example, greater values of $\lambda\sigma$ yield better resolution in the time domain, while smaller values yield improved resolution in the frequency domain, where the frequency and time resolution satisfy a Heisenberg-like uncertainty relation~\cite{Press2007,Cohen2014}. Windowed FFT of the N\'eel vector produces  $N^z(\omega,t_c)$, whose power spectrum in Figs.~\ref{fig:fig2}(b) and \ref{fig:fig2}(d) reveals excitation of the so-called ``bright magnon''~\footnote[2]{The jargon ``bright magnon'' was introduced in experiments on CrSBr~\cite{Bae2022,Diederich2022,Diederich2025,Sun2024}. For a clear explanation of the distinction between optically bright and dark magnon modes, see Ref.~\cite{Diederich2022}.} at the frequency $\omega_b$. In the presence of nonzero $U$ and thereby generated excitons, we find a longer lifetime of excited magnons in Fig.~\ref{fig:fig2}(d). Interestingly, both Figs.~\ref{fig:fig2}(b) and ~\ref{fig:fig2}(d) show short (for $U=0$) vs. longer (for $U\neq 0$) living magnons, respectively, that are excited at frequencies {\em below}  $\omega_b$. We, therefore, provide additional analysis in Appendix~\ref{sec:appendix} confirming [Fig.~\ref{fig:figS1}(a)] that such magnons are a direct consequence of $J_{K} \neq 0$. Thus,  our TDNEGF+LLG+EX framework explains microscopically the same ``bright magnon'' observed in  experiments on CrSBr~\cite{Bae2022,Diederich2022,Diederich2025,Sun2024}, using fsLP whose central frequency is above-gap, as the consequence of time-dependent STT [Eq.~\eqref{eq:stt}] mediated by photoexcited unbound electron-hole pairs or both them and excitons.

Furthermore, the excited magnons will  introduce the second nonequilibrium  drive  into the subsystem of electrons. Such a time-dependent drive  can lead to {\em pumping}~\cite{Evelt2017,Suresh2020,Ciccarelli2015,Kapelrud2013} of electronic spin and charge currents. They can be differentiated from currents pumped~\cite{Bajpai2019} by fsLP (as the first nonequilibrium drive) by their spectral content, as fsLP and magnons have vastly different frequencies.   We compute  charge current $I_R(t)$ [Figs.~\ref{fig:fig4}(a) and \ref{fig:fig4}(c)] pumped into the right NM lead, and perform windowed FFT on it to obtain  $|I_R(t_c,\omega)|^2$  plotted [Figs.~\ref{fig:fig4}(b) and \ref{fig:fig4}(d)] in the same frequency range where magnons are found in Fig.~\ref{fig:fig2}. The power spectrum $|I_R(t_c,\omega)|^2$ exhibits a peak at $\omega_b$, thus, by attaching an additional external circuit to recent experiments and by analyzing pumped current into that circuit, one could confirm optical excitations of magnons. Interestingly, pumped current could also sense the presence of excitons---in $U \neq 0$ case long-lived high harmonic generation (HHG) emerges in Fig.~\ref{fig:fig4}(d),  particularly at frequency $4 \omega_b$.    

Finally, we examine EM radiation emitted by time-dependent  local (or bond~\cite{Nikolic2006})  charge currents $I_{ia\rightarrow jb}(t)$ [Eq.~\eqref{eq:bondcharge}] within AF semiconductor CAR in Fig.~\ref{fig:fig0}. Note that EM radiation emitted by ultrafast-light-driven magnetic materials and their heterostructures is routinely  used~\cite{Beaurepaire2004,Seifert2016,Wu2017,Seifert2023} in spintronic experiments as a probe of coupled spin-charge dynamics~\cite{Kefayati2024,Kefayati2025,VarelaManjarres2024,Francesco2025} of such far-from-equilibrium systems. We compute the $x$-component $E^x_{\mathrm{FF}}(\mathbf{r},t)$ of the electric field  of EM radiation in the far-field (FF) region via the Jefimenko~\cite{Jefimenko1966,McDonald1997} formula  using   $I_{ia\rightarrow jb}(t)$  as the source~\cite{Ridley2021,Suresh2023}. The FF region is defined by radiation decaying as $\sim 1/r$, so we use  $\mathbf{r}_0=(5a,0,1000a)$ as the observation point in the Jefimenko formula [Eq.~\eqref{eq:efield}]. The windowed FFT of $E^x_{\mathrm{FF}}(\mathbf{r}_0,t)$ in Figs.~\ref{fig:fig3}(a) and \ref{fig:fig3}(c)  yields $E^x_{\mathrm{FF}}(\mathbf{r}_0,t_c,\omega)$ whose power spectrum is plotted in Figs.~\ref{fig:fig3}(b) and \ref{fig:fig3}(d), respectively.
Similarly to $|I_R(t_c,\omega)|^2$, the power spectrum  $|E^x_{\rm FF}(t_c,\omega)|^2$  contains an imprint of excited ``bright magnon'' at $\omega_b$ frequency, as well as HHG at $4\omega_b$  [Fig.~\ref{fig:fig3}(d)] signaling the presence of excitons and their role in STT exciting magnons in Fig.~\ref{fig:fig2}. Note that the frequency content of EM radiation by time-dependent current generated by excited magnons will be in the range of: THz in Fig.~\ref{fig:fig3} due to parameters selected in our quantum and classical Hamiltonians for convenience (i.e., to ensure we can complete time evolution within \mbox{$\sim 10$ ps}); sub-THz for realistic CrI$_3$ in Ref.~\cite{Zhang2020a}; and  GHz for realistic CrSBr in Ref.~\cite{Diederich2025}. 
 
\section{Conclusions and Outlook}\label{sec:conclusions}

\subsection{Conclusions}

Using  time-dependent quantum transport~\cite{Gaury2014,Popescu2016} combined~\cite{Bajpai2019a,Petrovic2018} with the classical LLG equation~\cite{Evans2014,Teuling2025}, as well as extended~\cite{Cistaro2022,Murakami2020} to include Coulomb-binding of photoexcited electrons and holes into excitons, we explain that the microscopic mechanism behind recent experimental observations of optical excitation of coherent~\cite{Pirro2021,De2024} magnons in  2D AF semiconductors~\cite{Bae2022,Diederich2022,Diederich2025,Zhang2020a,Belvin2021,Sun2024,Brennan2024} or their vdW heterostructures~\cite{Zhang2020a} is a  nonequilibrium spintronic effect of spin torque type~\cite{Ralph2008,Locatelli2014}. The {\em strongest experimental support} for our theory comes from the fact that without applying a magnetic field to induce sufficient {\em noncollinearity}  [i.e., their canting in Fig.~\ref{fig:fig0}]  between LMMs of two layers of CrSBr,  no magnon excitation is detected in Refs.~\cite{Bae2022,Diederich2022,Diederich2025}. This is because for collinear or nearly collinear (such as due to thermal fluctuations~\cite{Suresh2023,Ghosh2022}) LMMs, STT in Eq.~\eqref{eq:stt} is vanishingly small. Thus, our microscopic theory displaces the need for a phenomenological  ``impulsive perturbation''~\cite{Diederich2025,Zhang2020a} typically invoked within the classical LLG equation alone (i.e., without considering photoexcited electrons), as a standard tool employed to interpret recent experiments. It also connects ultrafast carrier dynamics in 2D magnetic materials to conventional spintronics, where STT is a cornerstone effect but is usually studied~\cite{Ralph2008,Locatelli2014,Nikolic2018,Belashchenko2019} in magnetic heterostructures driven by a small DC bias voltage or current pulses. Finally, it predicts additional charge pumping  by excited magnons (discussed in prior  spintronic literature~\cite{Ciccarelli2015,Suresh2020,Kapelrud2013,Evelt2017}, but for magnetic systems driven by GHz fields rather than light), as well as signatures of excitons in high harmonics~\cite{VarelaManjarres2023}  [Figs.~\ref{fig:fig4} and ~\ref{fig:fig3}] of pumped charge current. Thus, such unexplored signals of excited magnons and their interaction with excitons could be exploited in future experiments. Regarding which signal to choose for future experiments, both detection of pumped charge current in electrical contacts or its EM radiation contain nearly identical information  [compare Fig.~\ref{fig:fig4} vs. ~\ref{fig:fig3}]. So, the choice will largely depend on how easy it is to fabricate an optoelectronic device where 2D material is attached to electrical contacts. If that is possible (see, e.g., device in Fig.~1 of  Ref.~\cite{McIver2019}), and pumped current has spectral content in the GHz or smaller frequency range; we recommend analyzing such current flowing into an external attached circuit. If not, one can detect far-field EM radiation in the GHz or THz range of frequencies, which is a well-established experimental technique~\cite{Beaurepaire2004,Seifert2016,Seifert2023,Wu2017} in ultrafast spintronics when detection of excited spin and charge currents via an external circuit is cumbersome or impossible.

\subsection{Outlook}

Another recently developed quantum-classical formalism~\cite{Kudlis2023}---where the Schr\"{o}dinger equation, employing the first-principles derived effective Hamiltonian in terms of composite bosons
describing excitons, is coupled self-consistently to the classical LLG equation for LMMs---has found excitation of magnons and total magnetization reversal in ultrafast light-driven CrI$_3$. The same approach could, in principle, be applied to other 2D AF semiconductors. However, it does not consider unbound electron-hole pairs, which coexist~\cite{Meineke2024} with excitons when using above-gap light. For example, while experiments on NiPS$_3$ have observed magnon excitation for subgap light tuned to exciton energy and, therefore, mediated by it, in the case of CrSBr, no magnons are detected for subgap light~\cite{Bae2025C}. This suggests  that considering {\em both} unbound electron-hole pairs and excitons is {\em required} to understand above-gap light excitation of CrSBr~\cite{Meineke2024}, as naturally captured~\cite{Perfetto2022,Perfetto2023,Cistaro2022} by the TDNEGF formalism but absent in wavefunction-based approach of Ref.~\cite{Kudlis2023}. Our TDNEGF+LLG+EX formalism [Fig.~\ref{fig:fig1}] formulated for open quantum systems [such as the one in Fig.~\ref{fig:fig0}] also naturally  computes spin and charge currents [Fig.~\ref{fig:fig3}] pumped out of the system or STT within the system (see movie provided as the SM~\footnotemark[1]), which cannot be studied by time-evolving wavefunction of closed quantum systems.

Specifically for CrSBr, the very recent experiments have also observed additional effects associated with exciton-magnon coupling, not found in other 2D AF semiconductors~\cite{Brennan2024}, where either magnons modulate exciton energy (by a few meV)~\cite{Bae2022}; or  HHG~\cite{Diederich2025} of magnons are generated at frequencies $n\omega_b$, with $n$ reaching surprisingly large values $n \gtrsim 20$. Regarding the former, our self-consistent loop in Fig.~\ref{fig:fig1}, where magnon presence adds a time-dependent term into the electronic Hamiltonian, could explore this effect by focusing on exciton properties, rather than on magnon properties in the focus of the present study. Regarding the latter, our theory reproduces only ``bright magnon'' at frequency $\omega_b$, while not  capturing [Fig.~\ref{fig:fig2}]  HHG  of magnons at $n\omega_b$ frequencies.  Since the LLG equation is nonlinear, it can in principle describe~\cite{GarciaGaitan2025,Zheng2023} 
magnon-magnon interactions as one of the key ingredients~\cite{Huang2024} for large $n$. Such interactions are enhanced~\cite{Zhitomirsky2013} by experimentally induced noncollinearity (i.e., canting in Fig.~\ref{fig:fig0}) of LMMs within CrSBr~\cite{Diederich2025}. Although Fig.~\ref{fig:fig2} suggest that LLG description is  insufficient to capture the HHG of magnons in CrSBr,  we offer a hint in the Appendix [Fig.~\ref{fig:figS1}(c)] toward possible ingredients required to describe this effect---there we find that increasing $J_{K}$ to an unrealistically large value (i.e., ten times larger than used in Figs.~\ref{fig:fig2}--\ref{fig:fig3}, as a parameter whose precise value is difficult to extract from first-principles calculations~\cite{Kudlis2023}) does yield HHG, but at noninteger values $n$. 

\begin{figure*}
		\centering
		\includegraphics[scale=0.32]{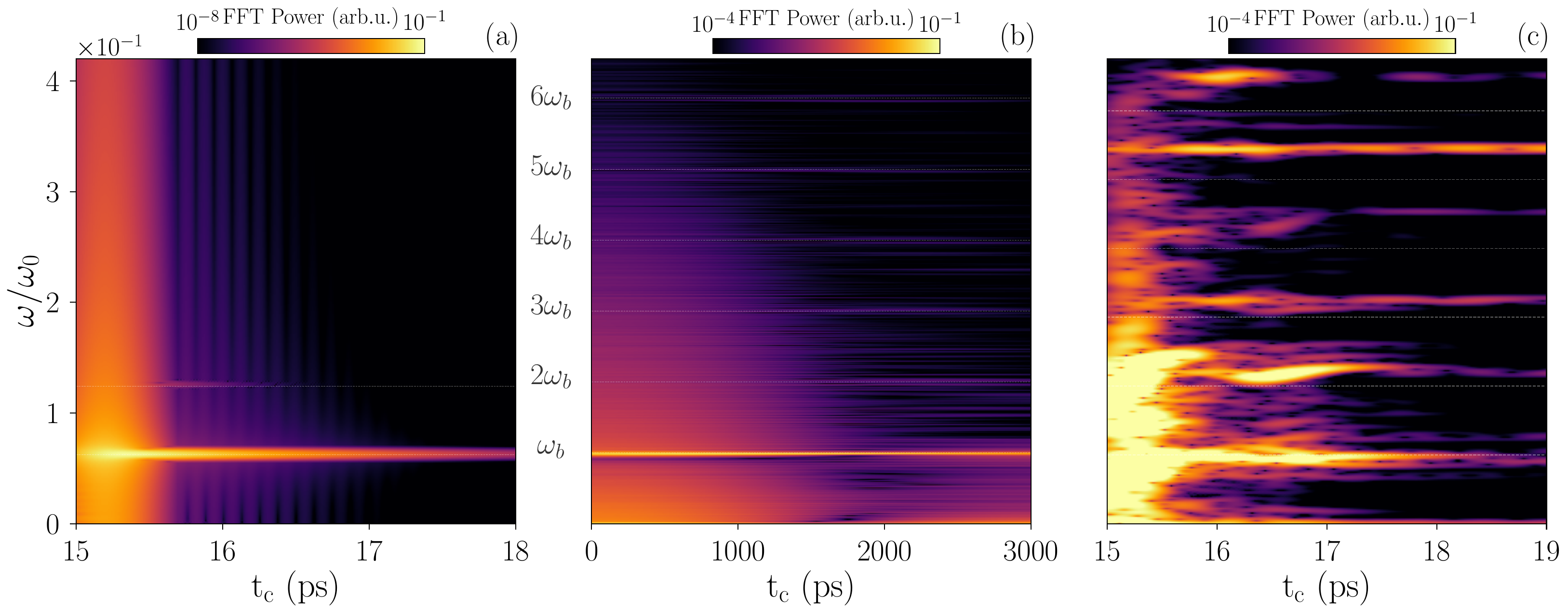}
		\caption{Power spectrum of windowed FFT~\cite{Press2007,Cohen2014},  $|N^z(t_c,\omega)|^2$, of   
        the $z$-component of the N\'{e}el vector [Eq.~\eqref{eq:Neel_vec}] for: (a) and (c) TDNEGF+LLG-computed dynamics in the absence of excitons ($U=0$); or (b) the same dynamics probed experimentally in CrSBr at externally applied magnetic field \mbox{$B=0.3$ T} [note that the  same data from panel (b) was analyzed in Fig.~4b of experimental Ref.~\cite{Diederich2025} but using standard FFT]. The dynamics of the N\'{e}el vector in panel (a) is initiated by photoexcited electrons, i.e., in the same fashion as in Fig.~\ref{fig:fig2}, but then for \mbox{$t \ge 15$ ps} we switch off  ($J_{K}=0$)  Kondo exchange interaction between flowing electronic spins and LMMs. In panel (c), we increase  the value of $J_{K}$ used in Fig.~\ref{fig:fig2} by ten times. Note that $\omega_b$ is larger in our calculations within panels (a) and (c) than in the experimental data of panel (b), so for easy comparison we rescaled the ordinate of panel (b).} \label{fig:figS1}
\end{figure*}

We anticipate that replacing the LLG equation with semiclassical theory but capable of capturing nonlinear dynamics that includes nonclassical effects, as proposed very recently in Ref.~\cite{Koerber2025} and applied to antiferromagnets, and/or by treating excitons beyond tMFT could explain the experimentally observed~\cite{Diederich2025} HHG of magnons and the role of many-body interaction~\cite{Zhang2025} with excitons in this process. For example, within the TDNEGF formalism, we can include additional self-energies to surpass the tMFT description of excitons,  as well as systematically improve them~\cite{Murakami2020,stefanucci2025,Perfetto2022,Perfetto2023}. Note 
that tMFT, or different versions of 
diagrammatic approximations for the self-energy in TDNEGF formalism, have been compared with the exact benchmark obtained from tensor network-based algorithms in Ref.~\cite{Murakami2020}, finding that the most important difference between the tMFT and the rest is the lack of proper damping of the induced coherent oscillations. We relegate possible extensions (as delineated in this paragraph) of the TDNEGF+LLG+EX  framework to future studies.

\begin{acknowledgments}
We thank Y. J. Bae for immensely valuable insights regarding  published and unpublished experimental data. J. V.-M. and Y. R. were supported by the University of Delaware (UD) Research Foundation Strategic Initiative Award. Y. R. was also supported by the US National Science Foundation (NSF) through the UD Materials Research Science and Engineering Center (MRSEC), DMR-2011824. B. K. N. was supported by the US Department of Energy (DOE) Grant No. DE-SC0026068. 
\end{acknowledgments}

\appendix 

\section{Additional examples of windowed FFT related to Fig.~\ref{fig:fig2}}~\label{sec:appendix}

An additional Fig.~\ref{fig:figS1} applies windowed FFT to two cases with artificially switched off or increased $J_K$ parameter, as well as to  experimental data of Ref.~\cite{Diederich2025}.  This then helps to clarify our results in Fig.~\ref{fig:fig2}.  Figure~\ref{fig:figS1}(a) analyzes the effect of photoexcited electrons---for simplicity, without their binding into excitons [$U = 0$ in Eq.~\eqref{eq:coulomb}]---on the time evolution of optically excited magnons. In particular, the goal is to understand the origin of magnon peaks in Figs.~\ref{fig:fig2}(b) and ~\ref{fig:fig2}(d) below the frequency $\omega_b$ of ``bright magnon.'' For this purpose, we allow photoexcited electrons to ignite magnons microscopically (instead of adding phenomenological ``impulsive 
perturbation''~\cite{Diederich2025,Zhang2020a} into the LLG equation) via  STT exerted by electronic spin current~\cite{Ralph2008,Nikolic2018,Belashchenko2019}, but then we immediately switch off  Kondo exchange $J_K$  interaction between flowing electronic spins and LMMs. In other words, we use the full self-consistent loop  [Fig.~\ref{fig:fig1}] in the TDNEGF+LLG  framework before $t=15$ ps in Fig.~\ref{fig:figS1}(a), and then for \mbox{$t \ge 15$ ps} $J_K(t)=0$, so that the TDNEGF and LLG parts of the loop are disconnected.  Such calculations lead to a longer lifetime of magnon at frequency $\omega_b$~\cite{Diederich2025} and its second harmonic $2\omega_b$ in Fig.~\ref{fig:figS1}(a)---compare the length of bright lines in Fig.~\ref{fig:figS1}(a) at these two  frequencies with their counterparts from Fig.~\ref{fig:fig2}(b). The longer lifetime is the consequence of removing {\em nonlocal} damping~\cite{Bajpai2019a,ReyesOsorio2024,Reyes2024,Sayad2015,Zhang2009,Yuan2016,Verba2018} by $J_{K}=0$ for \mbox{$t \ge 15$ ps}, which is otherwise  provided by the fermionic bath of electrons within TDNEGF calculations of the TDNEGF+LLG  loop.  In contrast,  such damping is present in Fig.~\ref{fig:fig2} during the whole evolution time. Note that the counterpart of the magnon at frequency $2\omega_b$, as clearly visible in Fig.~\ref{fig:figS1}(a), is of vanishingly short lifetime in Fig.~\ref{fig:fig2}(b) due to nonlocal  damping being operative in the latter case. The remaining finite lifetime of the $2\omega_b$ magnon in Fig.~\ref{fig:figS1}(a) is due to retained conventional local Gilbert damping of strength $\alpha_G$ [Eq.~\eqref{eq:llg}]. Importantly,  switching off the Kondo exchange interaction {\em removes} [Fig.~\ref{fig:figS1}(a)]  bright lines of excited magnons that are otherwise present in Fig.~\ref{fig:fig2}(b) at frequencies below $\omega_b$.  This finding confirms that magnon  peaks below $\omega_b$ are due~\cite{ReyesOsorio2025} to photoexcited electrons, which are explicitly included in our TDNEGF+LLG framework but are absent from the LLG equation used on its own in Fig.~\ref{fig:figS1}(a) for \mbox{$t \ge 15$ ps.} 

For comparison, we apply windowed FFT~\cite{Press2007,Cohen2014} to experimental data from Ref.~\cite{Diederich2025} generated in the course of the detection of optically excited magnons in CrSBr. Such   analysis replicates in Fig.~\ref{fig:figS1}(b) ``bright magnon'' at frequency $\omega_b$, while additionally (in comparison with standard FFT performed in Ref.~\cite{Diederich2025}) revealing its long lifetime. We also unravel shorter-lived magnons at high harmonics $n\omega_b$. However, the power spectrum of windowed FFT is too noisy in  Fig.~\ref{fig:figS1}(b) for frequencies  below $\omega_b$ to allow us to extract additional magnon peaks expected [Figs.~\ref{fig:fig2}(b) and ~\ref{fig:fig2}(d)] to be excited  due to flowing  electronic spins interacting  with LMMs.  We speculate that magnon peaks below $\omega_b$ frequency could be present in experiments, but they are short-lived and, therefore, invisible to the standard FFT employed in Ref.~\cite{Diederich2025}. 

Finally, in contrast to Fig.~\ref{fig:figS1}(a),  where $J_{K}$ is abruptly reduced to zero for \mbox{$t \ge 15$ ps}, in Fig.~\ref{fig:figS1}(c) we keep it nonzero for all times while increasing its value  ten times (\mbox{$J_{K}=0.1$ eV}) when compared to the same Kondo exchange interaction between  flowing electronic spins and LMMs used in Fig.~\ref{fig:fig2}. This intervention yields HHG in Fig.~\ref{fig:figS1}(c), but at frequencies that are not exact integer multiples $n \omega_b$. While such a   finding does not match integer HHG observed  experimentally~\cite{Diederich2025} and reproduced in Fig.~\ref{fig:figS1}(b), it does offer a hint for future extensions of the TDNEGF+LLG+EX framework, as discussed in Sec.~\ref{sec:conclusions}.


\bibliography{references}

\end{document}